\documentclass{edp-conf}
\usepackage{graphicx}
\begin{document}

\TitreGlobal{SF2A 2002}

%%-----------------------------
%%      the top matter
%%-----------------------------
\title{Deriving the redshift of distant galaxies with Gamma-Ray Burst
transient edges}

\author{Le Floc'h, E.}
\author{Duc, P.-A.,}
\author{Mirabel, I.F.}\address{Service d'Astrophysique, CEA--Saclay, 91191 Gif-sur-Yvette}

\runningtitle{Gamma-Ray Burst host galaxies }

\setcounter{page}{237}
% Keep this line, even if the page will be settled afterwards..
\index{Le Floc'h, E..}
\index{Duc, P.A.}
\index{Mirabel, I.F.}
% Repeat the authors here, this will help to make the final index

\maketitle

\begin{abstract}
Cosmological Gamma-Ray Bursts (GRBs) offer a unique perspective to
probe the evolution of distant galaxies. We discuss one of the
multiple benefits of this approach, i.e.  the detection of transient
edges in the GRB prompt phase emission. These absorption features can
be used to directly derive the redshift of GRBs and their host
galaxies without the need of any optical spectroscopic follow-up.

\end{abstract}
%
%%-----------------------------
%%      your text
%%-----------------------------
\section{GRBs and their host: a new approach on galaxy evolution}

Gamma-ray bursts (GRBs) are now regarded as one of the most promising
tools to probe the star formation in the early Universe.  There is
indeed increasing evidence that the long GRBs originate from the core
collapse of short-lived massive stars and they are thus considered as
direct tracers of active starbursts. The interests of this approach
are numerous:

\begin{itemize}
\item{}GRBs are not attenuated by intervening columns of gas and dust,
and identically probe unobscured star-forming regions and dusty
starbursts. They can trace therefore the history of star formation in
the Universe without any bias due to dust obscuration;
\vspace{-.2cm}
\item{}GRBs are likely detectable up to very high redshift because
they are beamed into relativistic jets;
\vspace{-.2cm}
\item{}The spectroscopic redshift of GRBs and their host galaxies can
de directly obtained using the features observed in the spectra of
GRBs and their transient counterparts at longer wavelengths.
\end{itemize}

So far, three different techniques have been used to derive these
redshifts from the intrinsic emission of GRBs and their afterglows,
including the detection of 1) absorption features in the spectra of
GRB optical transients, 2) iron K emission lines in X-ray afterglows,
and 3) absorption transient edges in the GRB prompt phase emission.
We examine this third issue hereafter.

\section{Determination of GRB redshift using prompt phase transient features}
In the GRB\,990705 event, Amati et al. (2000) reported the discovery
of a transient absorption edge at $\sim$\,3.8~keV in the prompt X-ray
emission of this burst (see Fig.\,1). They interpreted this feature as
the GRB-intrinsic signature of an iron-enriched absorbing medium at
z\,$\sim$\,0.86\,+/--\,0.17.  With the intention of confirming this
result, we performed VLT spectroscopic observations of GRB\,990705
host galaxy and derived a redshift z=0.8424 (Fig.\,1, see also Le
Floc'h et al. 2002).

\begin{figure}[h]
   \centering
%  If you have a figure, remove the comments % before...
  \includegraphics[width=12cm]{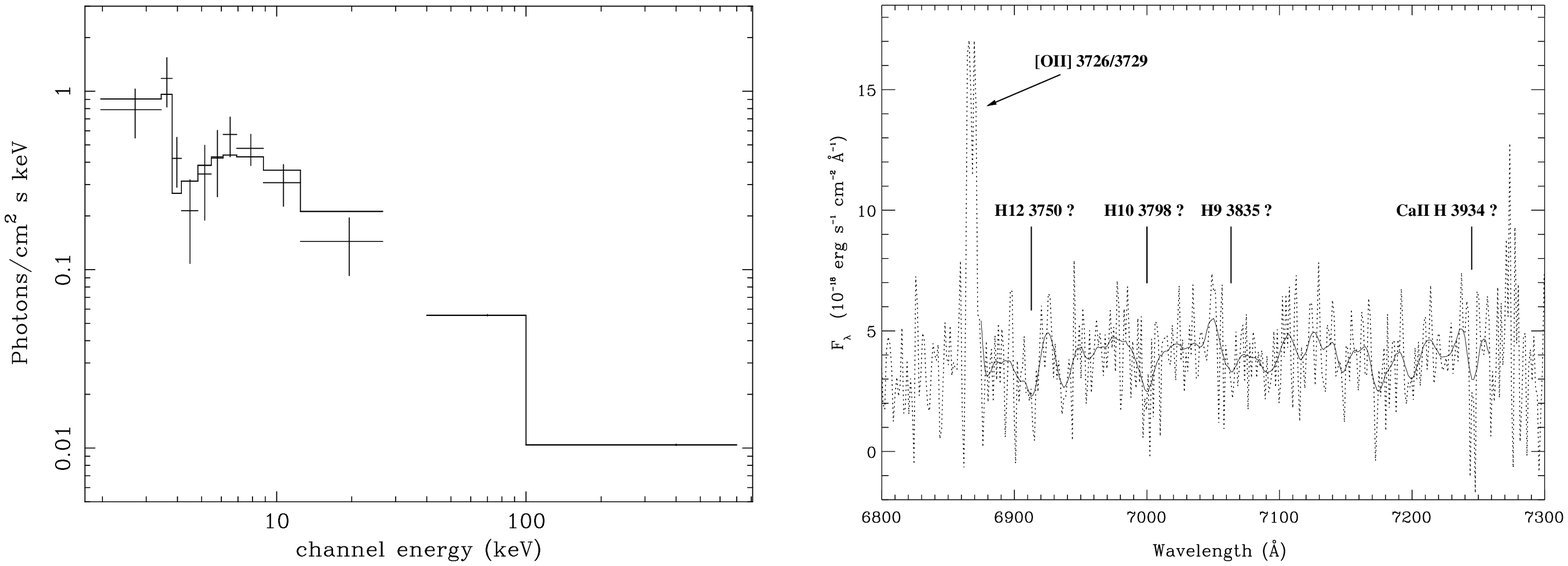}
      \caption{ {\it Left:\,} X-ray spectrum of GRB\,990705 with 
a transient edge at 3.8\,keV tracing an absorbing medium at z=0.86\,+/--\,0.17
(Amati et al. 2000). {\it Right:\,} Our VLT spectrum of GRB\,990705
host galaxy. The prominent [OII] doublet emission points to an unambiguous
redshift z=0.8424 for the burst and its host.}
       \label{figure_mafig}
   \end{figure}

This confirmation shows that intrinsic GRB properties, such as the
redshift and the physical conditions of the GRB--surrounding medium,
can be derived from the burst detection itself, without the need of
any afterglow to be detected and followed-up.  Eventhough GRB\,990705
is the only one burst in which such a transient edge has been observed
so far, which raises the question whether particular ionizing states
of the circum-burst environment are required to detect these
absorptions, this burst lies among the brightest GRBs ever detected
with the Beppo-SAX satellite.  This suggests that transient edges
could be a more common feature of GRB spectra, and highlights the
breakthrough that would be operated with the advent of more sensitive
X-ray detectors.

Future satellites indeed, such as the ECLAIRs experiment (Barret 2002)
could be entirely dedicated for studying the GRB prompt emission, and
may provide a systematic detection of these absorption lines.

%%-----------------------------
%%      your bibliography
%%-----------------------------

\end{document}